\documentclass[prl,aps,twocolumn]{revtex4-2}

\usepackage{amsmath}
\usepackage{amsfonts}
\usepackage{amssymb}

\usepackage{graphicx}
\usepackage{mathtools}

\usepackage{array}
\usepackage[colorlinks=true,citecolor=blue,linkcolor=blue]{hyperref}
\usepackage{mathrsfs}

\newcommand{\LNO}{La$_3$Ni$_2$O$_7$ }

\begin{document}

\title{Minimal two band model and experimental proposals to distinguish pairing mechanisms of the high-$\text{T}_c$ superconductor \LNO}

\author{Zheng-Duo Fan}
\affiliation{Harvard University, Cambridge, Massachusetts}

\author{Ashvin Vishwanath}
\affiliation{Harvard University, Cambridge, Massachusetts}

\begin{abstract}
The discovery of high-$\text{T}_c$ superconductivity in \LNO has opened the door to a new route to high‑temperature superconductivity, distinct from that in cuprates and iron‑based materials. Yet, despite intense recent activity, we lack experimentally testable protocols for distinguishing between different pairing scenarios. In this Letter, we construct a minimal two-band model that reproduces the Fermi-surface topology observed in recent ARPES measurements and DFT calculations, and we analyze superconductivity arising from two distinct pairing mechanisms. We show that these mechanisms yield sharply different responses to an applied perpendicular electric field. Thus, \LNO offers the unique opportunity to cleanly distinguish between different pairing scenarios.  Finally, we propose three concrete experimental proposals designed to distinguish these scenarios and thereby identify the pairing mechanism most relevant to the real material.

\end{abstract}

\maketitle

\section{Introduction}
In 2023, the bilayer nickelate \LNO  emerged as a striking new high $\text{T}_c$ platform, when superconductivity was discovered under pressure of about 14 GPa, with a $\text{T}_c$ as high as 80 K\cite{sun2023signatures,zhang2024high,mandyam2025uncovering}. A year later, superconductivity was also observed in thin film samples under ambient pressure \cite{ko2025signatures, zhou2412ambient,liu2025superconductivity,bhatt2025resolving}, with a superconducting temperature above 40 K. A recent review of this field is provided in \cite{wang2025recent}.

Early density functional theory (DFT) studies suggested that three bands cross the Fermi level, motivating theoretical efforts based on four orbital tight binding models \cite{luo2023bilayer}. However, after the superconductivity in the ambient pressure thin film \LNO was found, angle-resolved photoemission spectroscopy (ARPES) found only two Fermi surface sheets \cite{wang2025electronic}. Later DFT calculations were able to reproduce the ARPES data well by taking a more advanced exchange-correlation functional \cite{wang2024electronic}. In this Letter, we propose a {\em minimal two band model}, which captures the low energy physics and reproduces the Fermi surface in good agreement with the ARPES measurements and DFT calculations, offering a concise framework for future explorations. This is the first part of this work.

For understanding the pairing mechanism arising from the repulsive interaction $U$ in high $\text{T}_c$ superconductors, two paradigms have dominated discussions for decades — across cuprates, iron-based superconductor, and now nickelates: a strong correlated picture in which pairing is driven by superexchange interaction in the large $U$ regime, and a weak correlated picture in which pairing is mediated by spin and charge fluctuations in the small $U$ regime. Both  mechanisms have been notably successful in understanding (1) the pairing symmetry and (2) the relationship between the superconducting phase and the nearby antiferromagnetic ordered phase. However, which of these two pairing mechanisms is closer to describing real  materials — and even whether the distinction between them is sharply defined — remains under debate \cite{anderson2016last,scalapino2012common,lee2006doping}. Theoretical studies on \LNO can also be grouped into these two classes. Pairing from spin and charge fluctuation is studied by renormalization group \cite{gu2025effective,yang2023possible} and random phase approximation (RPA) \cite{liu2023s,liu2025origin,zhang2023trends,lechermann2023electronic}, while pairing from superexchange interactions is studied by \cite{yang2023interlayer,lu2024interlayer,xue2024magnetism,oh2025type}. The same question then reappears, which mechanism is closer to describing the real material? In this work, we show that \LNO offers a {\em unique opportunity} to address this long-standing question. Owing to its bilayer structure, the application of a perpendicular electric field provides a direct and tunable probe \cite{shao2411possible,huang2025effective}. We find that the two pairing mechanisms predict qualitatively different responses to such a field, offering experimentally testable signatures to determine which pairing mechanism is relevant to the real material. This is the second part of this work.

\section{Minimal Two-Band Model}
To reproduce the band structure obtained from DFT calculations, a four orbital tight binding model is usually required, consisting of two orbitals $d_{x^2-y^2}$ and $d_{z^2}$ in each of the two $\text{NiO}_2$ layers. Motivated by the recent ARPES and DFT results which show that only two Fermi-surface sheets exist near the Fermi level, we propose a minimal two band model to capture the low energy physics. 

\begin{figure}[ht]
    \centering
    \includegraphics[width=0.45\textwidth]{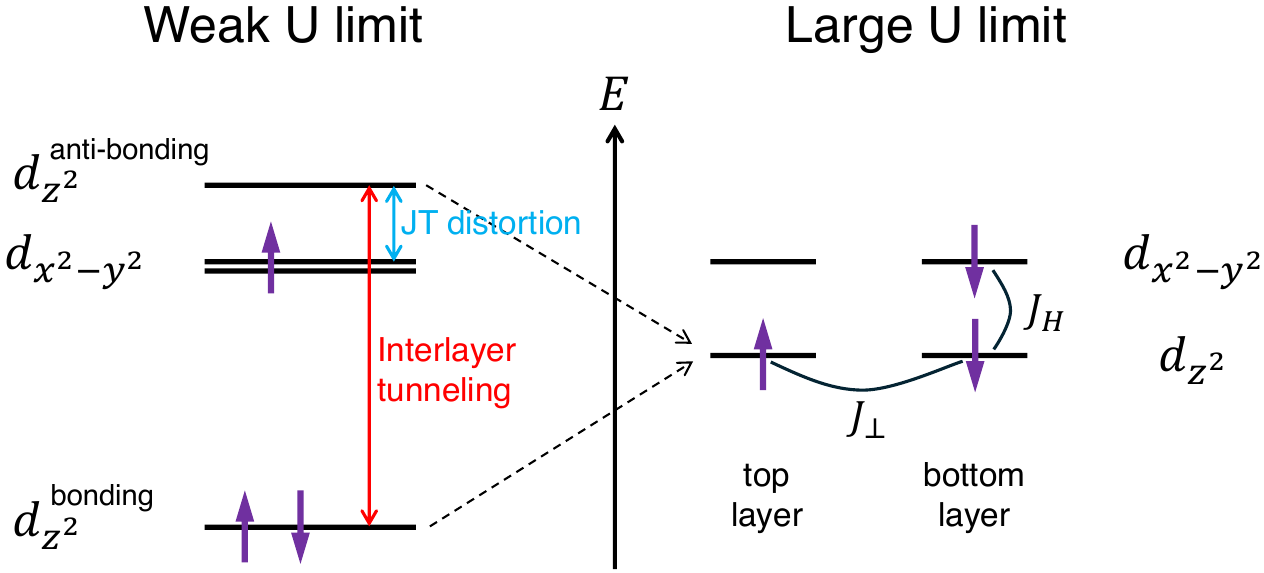}
    \caption{Electronic configuration of top layer and bottom layer atoms, vertically stacked above a single planar site.}
    \label{model}
\end{figure}

The model is constructed as follows. The $d_{z^2}$ orbital forms bonding and anti-bonding states due to inter-layer hopping. As shown by ARPES and DFT, this hopping is so strong that the bonding and anti-bonding states are far below and far above the Fermi energy, respectively (left panel of Fig. \ref{model}). The $d_{x^2-y^2}$ orbital dominates near the Fermi level, and the residual effect of the $d_{z^2}$ orbital is to mediate an effective interlayer hopping between $d_{x^2-y^2}$ electrons. Because direct $d_{x^2-y^2}$ - $d_{x^2-y^2}$ hopping between layers is weak, this second-order process mediated by the $d_{z^2}$ orbital becomes the dominant interlayer coupling. The $d_{z^2}$-mediated hopping proceeds as follows: a $d_{x^2-y^2}$ electron first hops to a nearest neighboring $d_{z^2}$ orbital within the same layer, then hops to the other layer's $d_{z^2}$ orbital, and finally returns to the $d_{x^2-y^2}$ orbital in that layer. This process gives rise to an interlayer hopping amplitude proportional to $(\text{cos}(k_x)-\text{cos}(k_y))^2$, reflecting the momentum dependence of the $d_{x^2-y^2}$ - $d_{z^2}$ hybridization.

For convenience, we introduce a two component field $\psi_k^\dagger=(c_{kt}^\dagger , \, c_{kb}^\dagger)$, where $c_{kl\alpha}^\dagger$ create the Ni-$d_{x^2-y^2}$ electron with wave vector $k$, layer $l=t,b$ and spin $\alpha=\uparrow, \downarrow$. The two-band minimal model is:
\begin{align}
&H=\sum_{k\alpha}\psi_{k\alpha}^\dagger \, ((\epsilon(k)-\mu)\text{I}+\eta(k)\sigma^x)\, \psi_{k\alpha} \label{tbmodel}  \\
&\epsilon(k)=-2t(\text{cos}(k_x)+\text{cos}(k_y))-4t'\text{cos}(k_x)\text{cos}(k_y)      \nonumber \\
&\eta(k)=-t_z(\text{cos}(k_x)-\text{cos}(k_y))^2    \nonumber
\end{align}

\begin{figure}[ht]
    \centering
    \includegraphics[width=0.45\textwidth]{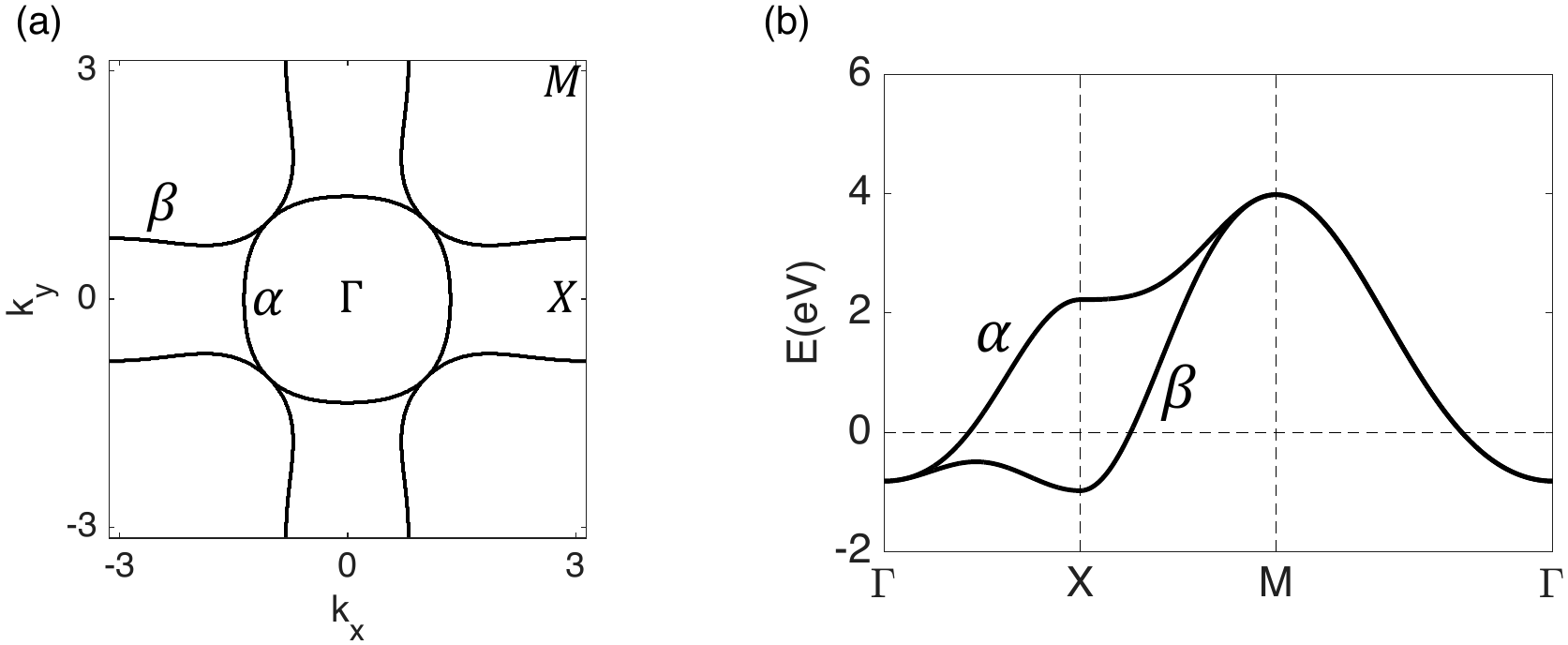}
    \caption{(a) Fermi surface of our minimal two band model, with  parameters (in eV)  $t=0.6, \, t'=-0.12, \, t_z=0.4, \,  \mu=-1.1$. (b) The band structure of the minimal two band model.}
    \label{fs}
\end{figure}

In Fig. \ref{fs} we show the band structure of the model for a specific choice of hopping parameters $t=0.6, \, t'=-0.12, \, t_z=0.4, \,  \mu=-1.1$, in unit of eV. We can see that this minimal two-band model captures essential features of the band structure: the $\alpha$ and $\beta$ bands are degenerate along $\Gamma \rightarrow \text{M}$ direction, where $\eta(k)=0$, consistent with the ARPES \cite{wang2025electronic,yang2024orbital} and DFT \cite{wang2024electronic} results.

In the following sections, we study the superconductivity after adding the onsite repulsive interaction,
\begin{eqnarray}
H_I= U\sum_{il}n_{il\uparrow}n_{il\downarrow}
\end{eqnarray}

We study it in two limits, $U\ll t$ and $U\gg t$, whose corresponding electronic configurations are illustrated in Fig. \ref{model}. The theories of pairing mechanisms are distinct in these two limits, and the question is whether these two mechanisms can be sharply distinguished experimentally. We find that the distinction becomes experimentally accessible because the two mechanisms predict qualitatively different responses to an applied perpendicular electric field,

\begin{eqnarray}
H_D= \frac{D}{2}\sum_{i\alpha} (n_{it\alpha}-n_{ib\alpha})
\end{eqnarray}

\begin{figure}[ht]
    \centering
    \includegraphics[width=0.45\textwidth]{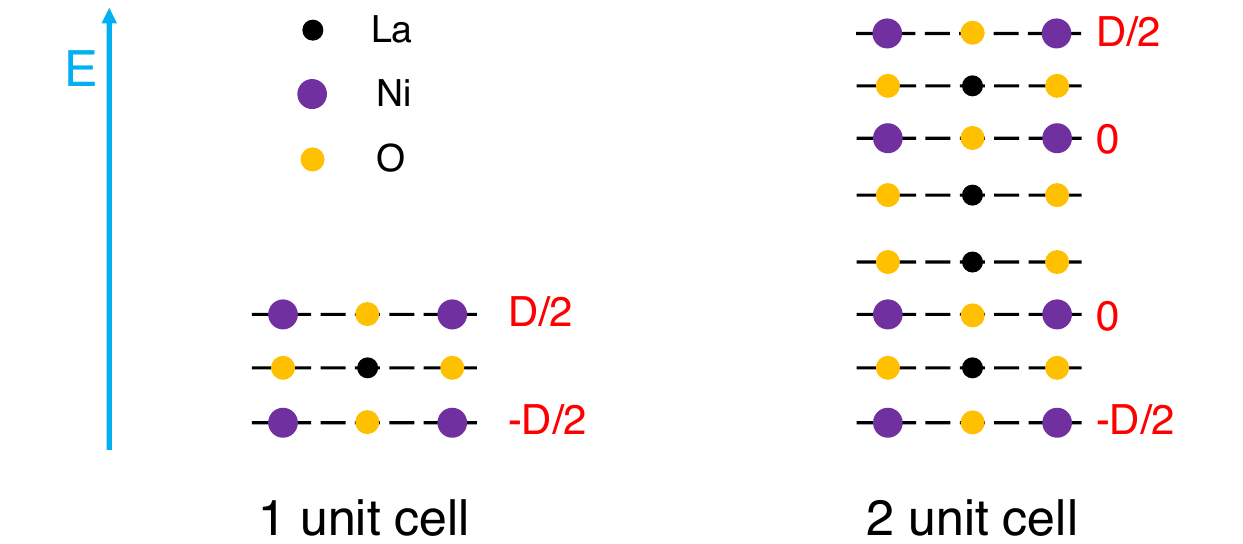}
    \caption{Schematic of the proposed setup with perpendicular electric field E applied to 1 unit cell and 2 unit cell films.}
    \label{setup}
\end{figure}

Due to the screening from the high electron densities in these metallic samples, the electric potential can only be applied on the  outer most $\text{NiO}_2$ plane \cite{ahn2003electric,bollinger2011superconductor}. Thus our calculations are directly applicable for the one unit cell and two unit cell thin film (Fig. \ref{setup}).

\section{RPA Calculation for Weak U Scenario}
When the onsite interaction $U$ is small, we investigate how superconductivity arises from spin and charge fluctuations\cite{scalapino1986d} within the weak interaction framework. So we first calculate spin and charge susceptibilities within the RPA approximation,
\begin{eqnarray}
&&\chi_s(k-k') = \chi_0(k-k')[1-U\chi_0(k-k')]^{-1}              \\ \nonumber
&&\chi_c(k-k') = \chi_0(k-k')[1+U\chi_0(k-k')]^{-1}
\end{eqnarray}
where $\chi_0(q)$ is the bare susceptibility,
\begin{eqnarray}
[\chi_0(q)]_{ab} =&& \frac{1}{N} \sum_{k,mn} \frac{n_F(\xi_n(k+q)) - n_F(\xi_m(k))}{\xi_m(k)-\xi_n(k+q)}  \\ \nonumber
&&\times a_n^a(k+q)a_n^b(k+q)^*a_m^b(k)a_m^a(k)^*
\end{eqnarray}
Here $\xi_{\pm} = \epsilon(k) \pm \sqrt{\eta(k)^2 + (\frac{D}{2})^2} - \mu$, $n_F$ is Fermi distribution function, and $a_n^a = \langle k, a|k, n\rangle$ is the eigen-wavefunction.

The singlet pairing interaction mediated by spin and charge fluctuation is given by \cite{scalapino1986d,graser2009near}:
\begin{eqnarray}
\Gamma_{ab}(k,k') = \frac{3}{2} U^2 [\chi_s(k-k')]_{ab} - \frac{1}{2} U^2 [\chi_c(k-k')]_{ab} + U
\end{eqnarray}

The dimensionless pairing strength $\lambda$ is calculated by solving the linearized gap equation,
\begin{eqnarray}
\lambda_\alpha\phi_\alpha(\hat{k}) = -\frac{1}{4\pi^2} \int_{FS} \frac{d\hat{k'}}{v_F(\hat{k'})} \, \Gamma_{\hat{k},\hat{k'}} \phi_\alpha(\hat{k'})
\end{eqnarray}
where $\hat{k}$ designates momentum on the Fermi surface, and $\Gamma_{\hat{k},\hat{k'}} = \sum_{a,b} \Gamma_{ab}(\hat{k},\hat{k'}) a_{n'}^a(\hat{k'}) a_{n'}^b(-\hat{k'}) a_{n}^a(\hat{k})^* a_{n}^b(-\hat{k})^*$. ($n$ and $n'$ are band indices of $\hat{k}$ and $\hat{k'}$, respectively)

\begin{figure}[ht]
    \centering
    \includegraphics[width=0.45\textwidth]{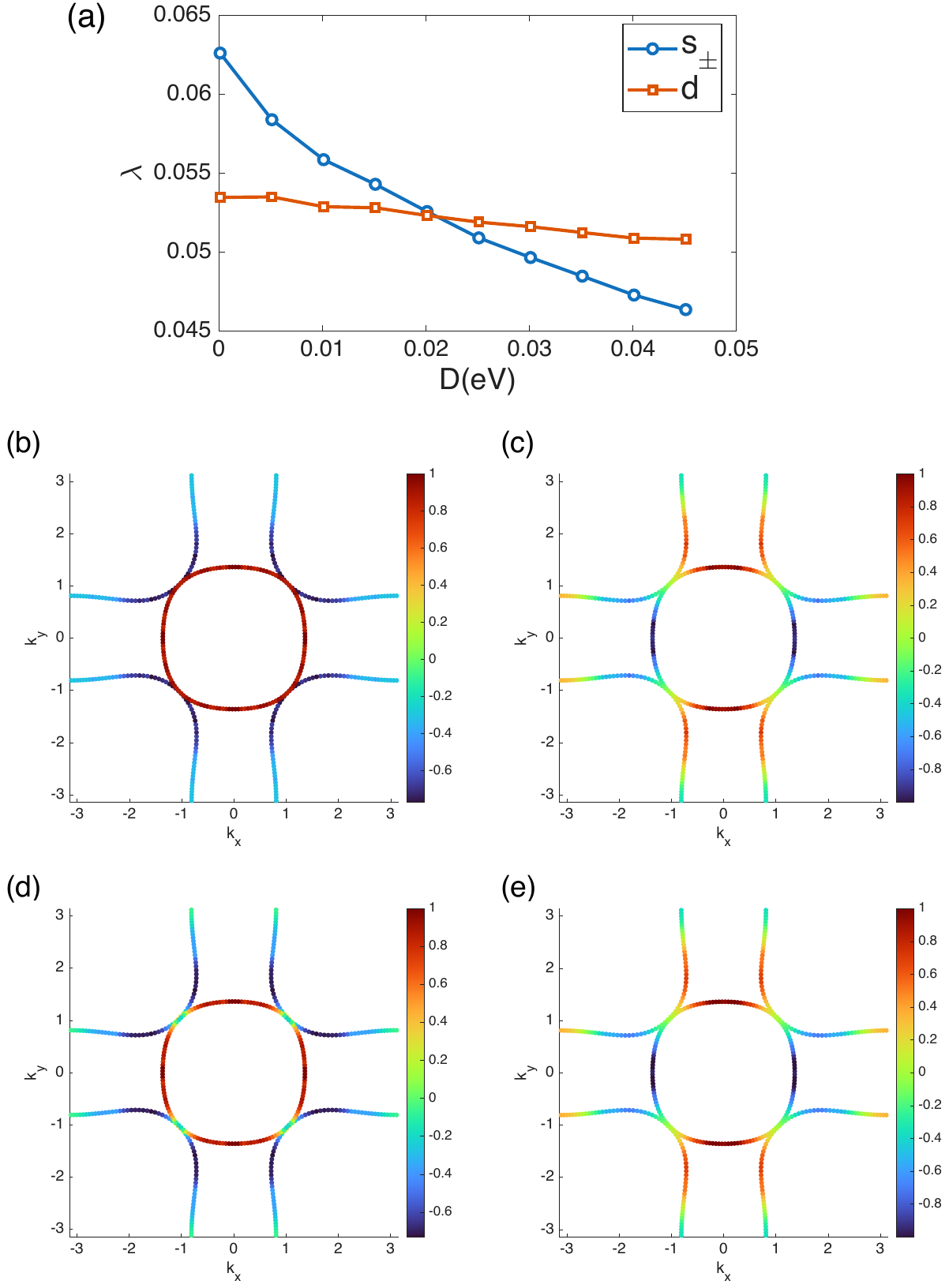}
    \caption{RPA calculation at U=1.6 eV. (a) Dimensionless pairing strength $\lambda$, which shows a phase transition from $s_\pm$ to $d$ wave on increasing $D$. (b) and (c) The gap functions of $s_\pm$ and $d$ waves at D=0 eV. (d) and (e) the gap function of $s_\pm$ and $d$ waves at D=0.02 eV.}
    \label{rpa}
\end{figure}

Our calculation of the dominant pairing channel as a function of the potential difference between layers $D$, shows a phase transition of the pairing symmetry from a  $s_{\pm}$ wave to a $d$ wave state with increasing $D$, as shown in Fig. \ref{rpa}(a). The gap values are plotted in Fig. \ref{rpa}(b)-(e).

The physical picture of the $s_{\pm}$ to $d$ transition is straightforward. Note that $s_{\pm}$ has a sign change between the two bands and is therefore favored by inter-band interaction. In contrast, the $d$ wave pairing changes sign within each band and is favored by the intra-band interaction. At $D$=0, the $\alpha$ and $\beta$ bands are anti-bonding and bonding states of the two layers, so the onsite repulsive interaction $U$ generates equally strong intra-band and inter-band scatterings. As a result, the $s_{\pm}$ and $d$ states are in competition, with the $s_{\pm}$ state slightly preferred. Applying a perpendicular electric field polarizes the layer weights — enhancing the top layer weight of the $\alpha$ band and the bottom layer weight of the $\beta$ band. This enhances the intra-band scattering, triggering a $s_{\pm} \rightarrow d$ transition at a critical field $D_c$.

To distinguish the $s_{\pm}$ and $d$ wave states in, for instance, ARPES experiments, we analyze their gap function shown in Fig. \ref{rpa}(b)-(e). For the $d$ wave gap function, it exhibits two sets of nodes: a symmetry enforced set of nodes along the Brillouin zone diagonal on both the $\alpha$ and $\beta$ bands, and the second set of nodes on the $\beta$ band near the Brillouin zone boundary ($\pi$, 0) and (0,$\pi$). In contrast, the $s_{\pm}$ gap function is nearly isotropic at $D$=0, but develops a pronounced minimum along the Brillouin diagonal as $D$ increases toward the phase transition point, although without forming an actual node. Because the diagonal minimum of the $s_{\pm}$ gap can become very small, it might be difficult to use the diagonal region to distinguish the two states. Instead, we propose that the additional node near the Brillouin zone boundary ($\pi$, 0) and (0, $\pi$) on the $\beta$ band — present only in the $d$ wave state — provides a sharp signature of the $s_{\pm} \rightarrow d$ transition. Given that we need to apply electric field on the sample, which limits the available experimental probes, we expect that recent developments including potassium surface dosing combined with a bottom gate architecture can open the possibility of using ARPES to directly investigate these predictions \cite{deng2025non}.

Our calculation gives $D_c\approx 0.02\text{eV}$. This number will change a lot if we use a different model parameter $t,t',t_z$, but the existence of the $s_{\pm} \rightarrow d$ phase transition and the underlying physical picture remain when the pairing is mediated by spin and charge fluctuations.

\section{$J_{\text{eff}}$ Model for Large U Scenario}
If onsite repulsion $U$ is greater than the bonding energy of $d_{z^2}$ electron, the $d_{z^2}$ electrons no longer form bonding and anti-bonding states, but instead form localized spins with an interlayer superexchange $J_\perp$ coupling (right panel of Fig. \ref{model}). This interaction of $d_{z^2}$ electrons can induce the same form of interaction on $d_{x^2-y^2}$ electrons by the Hund's coupling between two orbitals \cite{lu2024interlayer,xue2024magnetism,oh2025type,yang2025strong}. Here we adopt a simplified model that directly writes down the resulting effective interlayer exchange interaction $J_\text{eff}$ acing on the $d_{x^2-y^2}$ electrons. We believe that this model is enough to capture the essential physics relevant to the system's response to an applied electric field.

The model is,
\begin{align}
H=& -t \sum_{<i,j> l \alpha} c^{\dagger}_{il\alpha} c_{jl\alpha} -\mu\sum_{il\alpha}n_{il\alpha}    \nonumber \\
& + J_{\text{eff}} \sum_{i} \boldsymbol{S}_{it}  \cdot \boldsymbol{S}_{ib} + \frac{D}{2}\sum_{i\alpha} (n_{it\alpha}-n_{ib\alpha})   \label{J_eff}
\end{align}
where $c_{il\alpha}^\dagger$ create the Ni-$d_{x^2-y^2}$ electron at site $i$, layer $l=t,b$ and spin $\alpha=\uparrow, \downarrow$, $<i,j>$ is the nearest neighbor sites. $D$ is the potential difference between two layers by the applied electric field.

We ignore  interlayer tunneling in this model for the following reason. The interlayer tunneling mediated by $d_{z^2}$ orbital in Eq. \ref{tbmodel} is quenched since the $d_{z^2}$ electron forms a localized spin due to large $U$. For the direct interlayer tunneling of $d_{x^2-y^2}$ orbital, the relevant hopping path is Ni(top) - inner apical Oxygen - Ni(bottom), however, the overlap between $d_{x^2-y^2}$ orbital of Ni and $p_z$ orbital of Oxygen vanishes from wavefunction symmetry. The remaining symmetry allowed tunneling path involve longer distance through more sites, making them strongly suppressed. Although a small interlayer tunneling will frustrate pairing, calculations in \cite{oh2025type,lu2024interlayer} show that within a reasonable small value, this effect can be neglected.

The superexchange $J_{\text{eff}}$ plays the role of an attractive interaction. After decoupling it into the pairing channel, the above model is equivalent to the classic BCS Hamiltonian in a Zeeman field, where $\frac{3}{4}J_{\text{eff}}$ is the attractive pairing interaction $V$ in the BCS Hamiltonian, and $D/2$ acts as the Zeeman energy $\mu_BH$. The perpendicular electric field thus acts as a depairing field, suppressing the interlayer singlet pairing in the same manner that a Zeeman field suppresses superconductivity in a conventional BCS superconductor. The critical field of fully destroying superconductivity follows the Pauli (Chandrasekhar–Clogston) limit $D_c = 2\frac{\Delta_0}{\sqrt{2}}$ \cite{chandrasekhar1962note,clogston1962upper} ($\Delta_0$ is the BCS gap at $D$=$T$=0), and a finite momentum pairing state is expected in the phase diagram between the BCS state and the normal state \cite{fulde1964superconductivity,larkin1965inhomogeneous,kinnunen2018fulde}.

We calculate the phase diagram by mean field theory. The attractive interaction is decomposed by order parameter $\Delta(i) = V\langle f_{it\uparrow} f_{ib\downarrow} \rangle$ and we consider two mean field Ansatz:
\begin{eqnarray}
&&\Delta(\boldsymbol{r}) = \Delta e^{i2\boldsymbol{q}\cdot \boldsymbol{r}} \\ \nonumber
&&q=0    \qquad     \text{BCS Ansatz}    \\ \nonumber
&&q=(k_{F,b} - k_{F,t})/2    \qquad    \text{Fulde-Ferrell Ansatz}  
\end{eqnarray}

\begin{figure}[ht]
    \centering
    \includegraphics[width=0.45\textwidth]{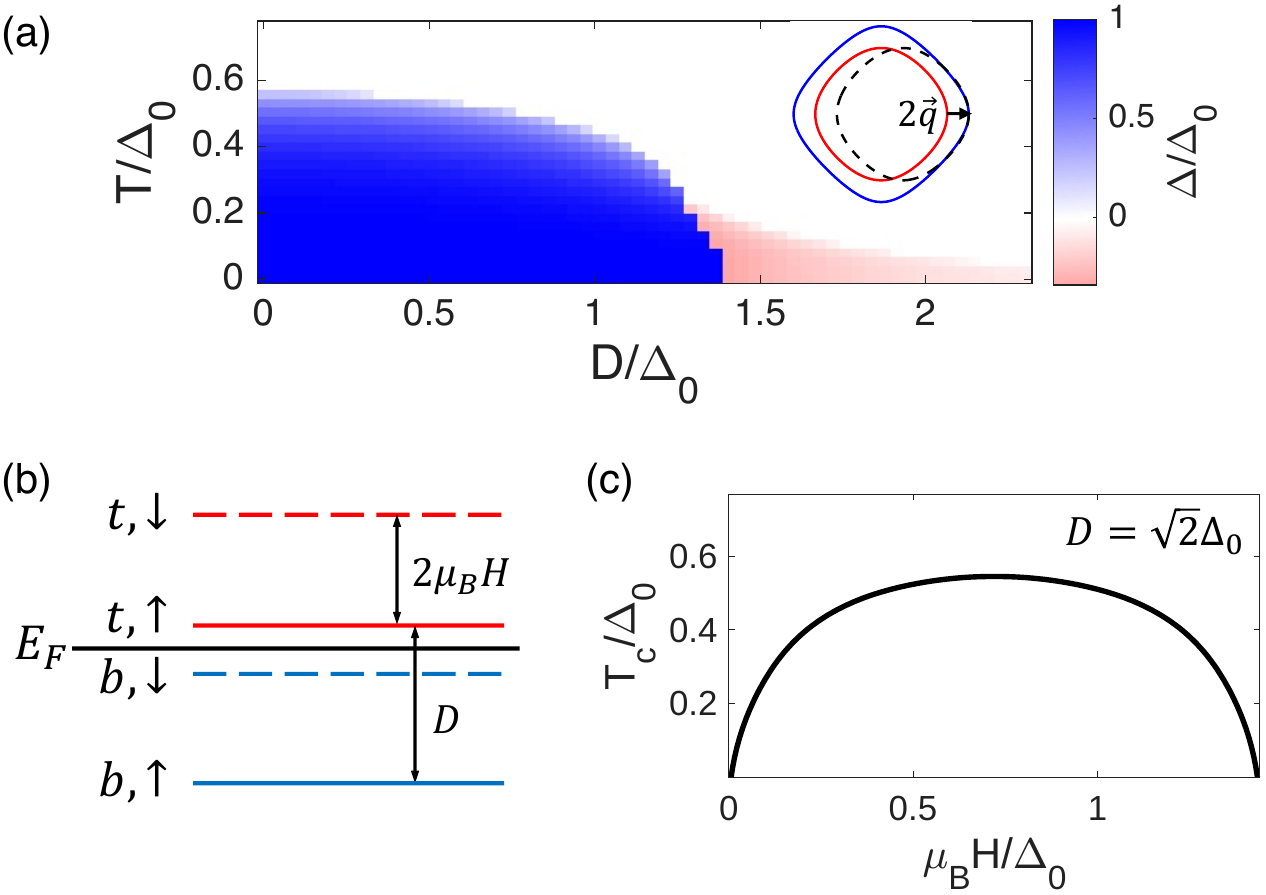}
    \caption{(a) Mean field phase diagram of the BCS and FF states. The gap value is encoded by color: $|\Delta_{\text{BCS}}|$ (blue) and $-|\Delta_{\text{FF}}|$ (red). t=V=1, $\mu=-1.4$ (so that filling is 1/4). The inset is an illustration of FF Ansatz. (b) Schematic bands shift after applying perpendicular electric field and parallel magnetic field. (c) Superconducting  $\text{T}_c$ as a function of parallel magnetic field for $D = 2\frac{\Delta_0}{\sqrt{2}}$, illustrating the compensation effect of the two fields.}
    \label{mf}
\end{figure}

The phase diagram is obtained by minimizing the free energy,
\begin{eqnarray}
&&F = -k_BT \sum_{k,\pm} \text{ln}(1+e^{-\beta(\pm E(k)+\xi_I(k))}) + \frac{\Delta^2}{V}            \\ \nonumber
&&E(k) = \sqrt{(\frac{\xi(k-q)+\xi(k+q)}{2})^2 + \Delta^2}     \\ \nonumber
&&\xi_I =  \frac{\xi(k-q)-\xi(k+q)}{2}  +  \frac{D}{2}
\end{eqnarray}
where $\xi(k)=-2t\text{cos}(k_x)-2t\text{cos}(k_y)-\mu$. The resulting phase diagram is shown in Fig. \ref{mf}(a).

Here we estimate where we can find the pair density wave phase. The known quantity is $\text{T}_{c,0}$, the superconducting transition temperature at $D=0$. According to the calculated phase diagram, the pair density wave state is expected to appear when the perpendicular field $D$ reaches $2.5\, k_B\text{T}_{c,0}$, and the temperature is below about $\frac{1}{3}\text{T}_{c,0}$. Taking $\text{T}_{c,0} = 40\text{K}$, the $D$ we need is 8.6 meV.

A further prediction arises when both a perpendicular electric field and an in-plane magnetic field are applied. Now we have two `Zeeman fields'. Since the former acts as a `Zeeman field' between layers and the latter as a conventional Zeeman field between spins, the two fields will have a compensation effect. The superconducting transition temperature $T_c$ takes its maximum value when these two fields match, $D = 2\mu_BH$, as shown schematically in Fig. \ref{mf}(b). The sharpest way to observe this compensation effect is the following: first tune the $D$ to the Pauli limit ($D = 2\frac{\Delta_0}{\sqrt{2}}$), where the superconductivity is completely destroyed. Then apply the parallel magnetic field. As the two Zeeman fields begin to compensate, $\text{T}_c$ will  {\em increase} and reach the maximum value $\text{T}_{c,0}$ (the $\text{T}_c$ in the absence of both electric and magnetic fields) at $\mu_BH = \Delta_0/\sqrt{2}$. This behavior is shown in Fig. \ref{mf}(c). Such a magnetic field induced reentrance of superconductivity would be a striking experimental signature.

\section{Conclusion}
In conclusion, we have described a minimal two-band model for \LNO, which well reproduces the topology of the ARPES and the DFT fermi surfaces. Exploiting the layer degree of freedom, we have shown that different theoretical pictures of superconductivity lead to experimentally distinguishable predictions when a perpendicular electric field is applied.

Within the weak $U$ framework, the applied field drives a transition of the superconducting gap symmetry from $s_{\pm}$ wave to $d$ wave. In contrast, in the large $U$ scenario, the same field induces a transition from a uniform BCS state to a pair density wave state, furthermore, there is a characteristic compensation effect on $\text{T}_c$ when an additional in-plane magnetic field is introduced. Although we discuss these two scenarios independently, the typical values of $D$ to see our predictions are $\sim$ 20 meV for the weak $U$ scenario and $\sim$ 10 meV for the large $U$ scenario. Thus, in fact, all the physics we discussed are of the same order of $D$. This fact sharpens our proposal for identifying these two pairing mechanisms. To realize this in the lab, one needs to fabricate a one or two unit cell thick \LNO film and integrate it into a dual‑gate device to generate a perpendicular electric field. The required field is on the order of $\sim \frac{50 \text{meV}}{d} = 0.1 \, \text{V/nm}$, where $d \sim 0.5 \text{nm}$ is the interlayer distance. Such field strength is readily accessible value by current experiments~\cite{zhang2009direct}.

We believe that these experiments will place strong constraints on the correlation regime relevant for the real material, thereby clarifying the pairing mechanism involved in \LNO. Beyond elucidating the physics of \LNO itself, such studies will also shed new light on how high superconducting $\text{T}_c$ are achieved in electronic materials.

Finally, our work also suggests a possible route to reach higher $\text{T}_c$ in thin film \LNO. Currently, film thickness must strike a balance.  If the film is too thick,  it no longer experiences enough epitaxial strain from the substrate. If it is too thin, we suspect that built in electric fields at the interfaces suppress superconductivity.   This trade off currently yields the highest $\text{T}_c$, 48 K, occurring in five unit cell films \cite{liu2025superconductivity}. For thinner films, we propose that an external electric field could compensate the built-in electric field and thereby stabilize — possibly even enhance — the superconducting $\text{T}_c$, since thinner films can accommodate a stronger substrate induced strain.

\section{Acknowledgments}
We would like to acknowledge Zhaoyu Han, Pavel A. Nosov, Tonghang Han, Julian May-Mann, Yahui Zhang, Fa Wang and Yi-Zhuang You for many helpful discussions. This work is supported by National Science Foundation grant NSF DMR-2220703 and by a Simons Investigator award (AV) which is a grant from the Simons Foundation.

\bibliographystyle{apsrev4-2}
\bibliography{refs}

\end{document}